\documentclass[twocolumn,amsmath,amssymb,pre,aps]{revtex4}   
\usepackage{graphicx}
\usepackage{dcolumn}
\usepackage{bm}
\usepackage{color}
\usepackage{graphicx,wrapfig,lipsum}
\usepackage{hyperref}

\renewcommand{\vec}[1]{\mathbf{#1}}
\newcommand{\squeezeup}{\hspace{-40.0mm}}

\newif\ifgraph

\graphtrue             %

\begin{document}
\title{Directed Autonomous Motion and Chiral Separation of Self-Propelled Janus Particles in Convection Roll Arrays}

\author{Poulami Bag\textsuperscript{a}, Shubhadip Nayak\textsuperscript{a}, Tanwi Debnath\textsuperscript{b}, and Pulak K. Ghosh\textsuperscript{a}\footnote[1]{Email: pulak.chem@presiuniv.ac.in}}

\affiliation{\textit{$^{a}$~Department of Chemistry, Presidency University, Kolkata 700073, India}}
\affiliation{\textit{$^{b}$~Department of Chemistry, University of Calcutta, Kolkata 700009, India}}

\date{\today}

\begin{abstract}
\begin{wrapfigure}{r}{5.0cm}
\squeezeup
\includegraphics[width=5.0cm, height=5.0cm]{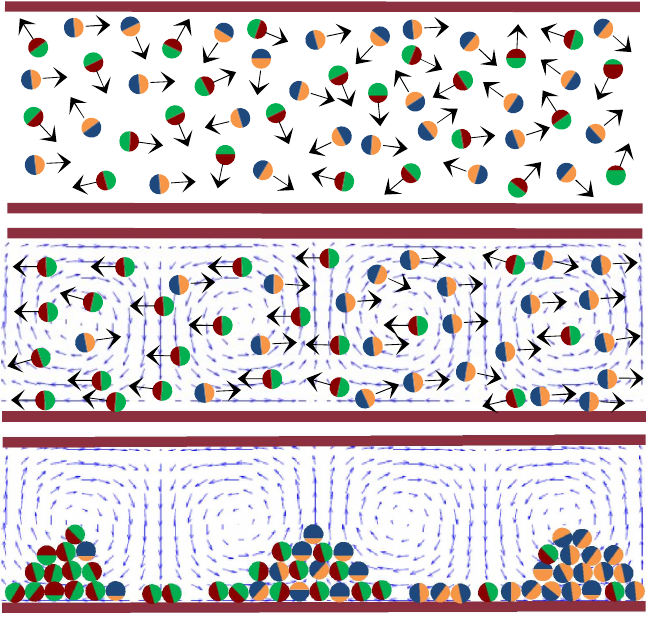}
\squeezeup
\end{wrapfigure}
 Self-propelled Janus particles exhibit autonomous motion thanks to engines of their own.   However,  due to randomly changing direction of such motion they are of little use for emerging nano-technological and bio-medical applications.  Here, we numerically show that  the motion of chiral active Janus  can be directed subjecting them to a  linear array of convection rolls. Rectification power of self-propulsion motion here can be made more than 60\%  which is much larger than earlier reports.  We show that rectification of chiral Janus particle's motion leads to conspicuous segregation of dextrogyre and levogyre active particles from a {\it racemic}  binary mixture.  Further, we demonstrate how efficiently the rectification effect can be exploited to separate dextrogyre and levogyre particles when their intrinsic torques are distributed with Gaussian statistics.
 
  
\end{abstract}


\maketitle

Artificial microswimmers are cleverly designed bio-inspired nanometer-to-micrometer sized self-propelled particles~\cite{Schweitzer-1,Schimansky-Geier-1,Ramaswamy,Medin,JPCL-1,Schauer,Magdanz,JPCL-2,Bunea,Srivastava,Granick,JPCL-3,Muller,Jiang}. 
They acquire self-propulsion taking advantage of local gradients resulting from some self-phoretic processes (e.g., diffusiophoresis, thermophoresis or electrophoresis) occurring at the particle-solvent interfaces~\cite{Bunea,Srivastava,Granick,Muller,Jiang}. 
As the self-propulsion acts as an external energy source, there is no balance between fluctuations and dissipations; thus, active particles (APs) are typical non-equilibrium systems and exhibit correlated Brownian motion with unusual transport features~\cite{Golestanian-1,Golestanian-2,Teeffelen,Stark,Volpe,Marchetti1,Marchetti2,Redner,Buttinoni}. According to Pierre Curie's conjecture, the motion of such particles can be rectified in spatial periodic structures with broken inversion symmetry. This effect is of concerted interest, both conceptual and technological.

Much effort is presently directed in establishing net transport of self-propelled particles in the absence of external bias~\cite{ourPRL,Reichhardt,Bao,Misko-1,Jaideep,Pietzonka,soft-sepa}. It has been proved that APs of the Janus kind exhibit directed autonomous motion in appropriate corrugated channels~\cite{ourPRL,our-chiral,cos}. Such an effect is very robust and rectification power is almost orders of magnitude larger than the conventional thermal ratchets~\cite{ourPRL}. Rectification of APs motion has further been examined in asymmetric periodic substrate potentials~\cite{Reichhardt,der}. These studies provide suggestive options in achieving reliable transport control of active Janus particles (AJPs) in periodic structures. However, unwanted particle-wall/substrate interactions pose challenges to observe autonomous motion in the desired way.   In this letter, we propose a more affordable option to establish ratchet transport of AJPs and sort them according to chirality\cite{technical}.

We numerically demonstrate that chiral AJPs exhibit directed motion in a linear periodic counter-rotating array of steady planer convection rolls. The rectification effect here does not require any additional ratchet potential. Diffusing heavy AJPs in convection roll arrays feel a periodic structure with broken upside-down symmetry due to their weight. This leads to rectification of chiral AP motion. It may appear that chirality~\cite{torq} (i.e., intrinsic torques) is an additional criterion for rectification. However, in practice, zero intrinsic torque of an AJP is very unlikely. Inevitable fabrication defects, even to a little extent [see Fig.1(c)],  produce  apparent asymmetry over the coated surface. Thus, an uneven rate of self-phoretic processes taking place over the active surface produces intrinsic torque in the self-propulsion motion.  

Our simulation results prove that the average drift velocity of autonomous motion can be made close to the self-propulsion speed by suitably adjusting advection speed.  We show how ratcheting can efficiently be utilized to separate dextrogyre ($d$) and  levogyre ($l$) AJPs from  mixtures (binary or multi-components).

\begin{figure}[tp]
\centering \includegraphics[width=9cm]{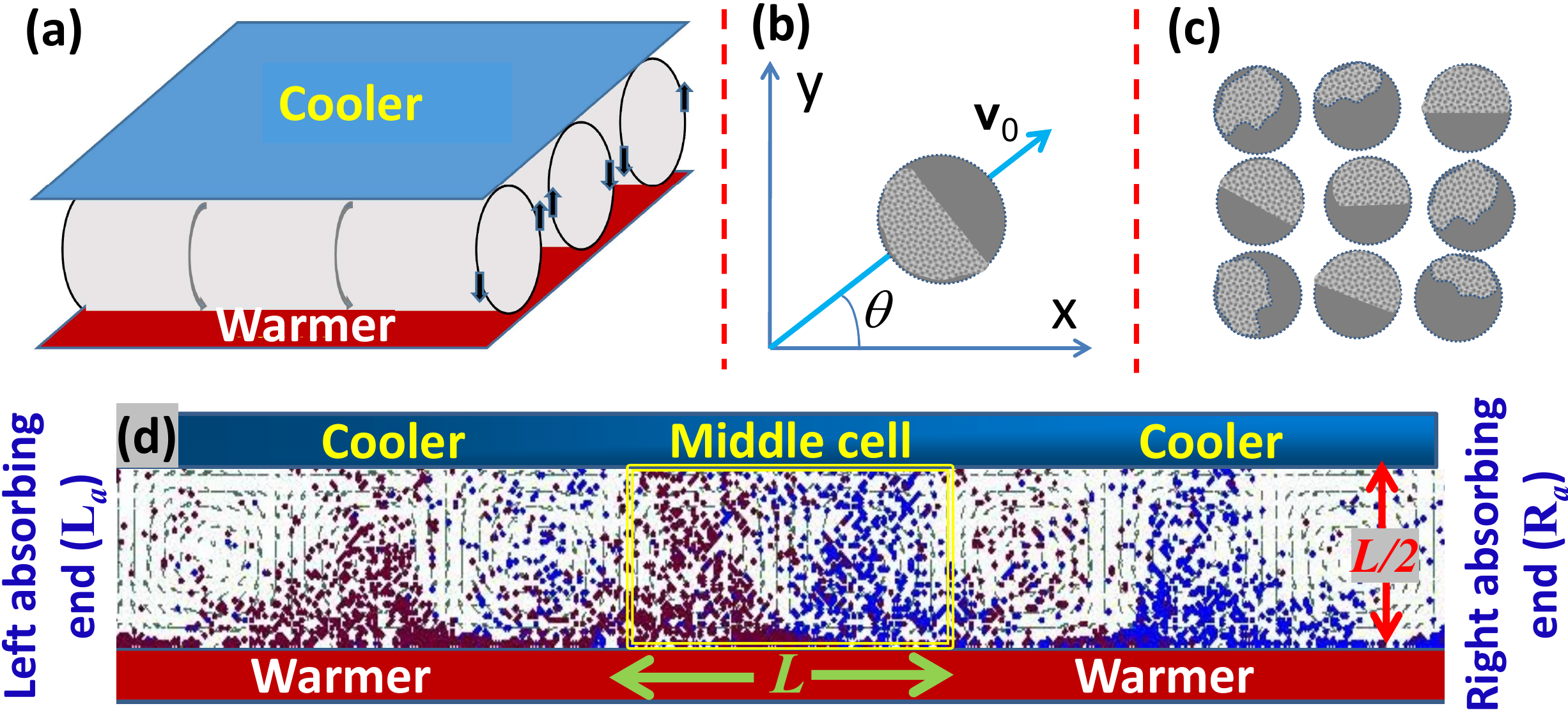}
\caption{ (Color online) (a) A sketch for practical realization of a linear convection roll array of Rayleigh-B\'enard type. The active particle of Janus kind depicting self-propulsion model encoded in Eq.~(\ref{LE}) is illustrated (b).(c) Schematic of Janus particles with fabrication defects, thus possessing intrinsic torques. (d) Chiral separation set up; at $t=0$ the mixture is injected in the middle cell  and particles are collected at the left absorbing ($L_a$) and right absorbing ($R_a$) ends. Distribution of $l$ (blue circle) and $d$  (wine circle)  AJPs in a convection array with stream function of Eq.(\ref{psif}) at $t=10^4$  starting with uniform distribution in the middle cells (framed in yellow). Parameters: $g=0.5,\; v_0 = 1, \;  D_0 =  D_\theta =0.01,\; |\Omega_0|=0.57 $.  }\label{F1}
\end{figure}

{ Note that many biomedical and nano-technological applications~\cite{ Bunea, Srivastava,soto,Minfeng,Zhiyong} require self-propelled particles with desired transport features. Further, dynamical properties of active particles largely depend on their intrinsic torque and hence chirality. As laboratory synthesis always produces AJP with different torques, it is desirable to sort them for practical purposes}.

We restrict our attention here to the low packing fraction region ($\phi \leq 0.1$) where impact of laminar flows around finite-size particles on advective drag of the suspension fluid can be ignored~\cite{Zoettl,Rusconi,Qi}. Further, our study focuses on low Reynolds number regimes where impact of hydrodynamics and inertia on the dynamics of nano- and micro-particles~\cite{Kirby}  can be ignored.


Model -- We consider a mixture of $N$ interacting  AJPs with different chirality diffusing in the convection array of stream function,
\begin{equation}
\label{psif}
\psi(x,y)= ({U_0L}/{2\pi})\sin({2\pi x}/{L})\sin({2\pi y}/{L}),
\end{equation}
The parallel walls at, $y=0$  and  $y=L/2$  act as  reflecting boundaries [see Fig.1]. The pair of counter-rotating rolls are characterized by the array unit cell of length $L$ and  the maximum advection velocity $U_0$. Combining these two constants, one can conveniently define the advection diffusion time scale as, $D_L=U_0L/2\pi$ and the roll vorticity,  $\Omega_L=2\pi U_0/L$.   

 A sketch of an ideal experimental set-up for convection rolls is presented in Fig. 1(a).
Stationary Rayleigh-B\'enard type counter-rotating cells can be  produced in a plane horizontal layer of fluid heated from the bottom~\cite{Bodenschatz,Getling}. Recent studies~\cite{wan1,wan2,wan3,hul} witness a significant advances in experimental methods controlling fluid dynamics. Considering the  Rayleigh-B\'enard type rolls are counter-rotating cylinders along the z-axis, the z coordinate of a suspended particle can be ignored; hence the reduced two-dimensional (2D) flow pattern of eq.~(\ref{psif}).

We consider AJPs as interacting disks of radius $r_0$.  They interact with each other through
a truncated Lennard-Jones (LJ) potential, $V_{ij} = 4\epsilon [\left({\sigma}/{r_{ij}}\right)^{12} -\left({\sigma}/{r_{ij}}\right)^{6}], \;\; {\rm if}\;\;  r_{ij} \leq r_m $; otherwise,  $V_{ij} =0 $ . Here, $r_m$ locates the potential minimum, and $\sigma = 2r_0$. We choose, $\epsilon = v_e\sigma^2/24$, with~\cite{Ran}, $v_e = v_0$ (self-propulsion velocity). 

The following equation encodes the dynamics of $i^{th}$ AJP in the mixture, 
\begin{eqnarray} \label{LE}
\dot {\vec r_{i}}= {\vec u}_{i} + \vec{g}+\sum_{j}\vec{F}_{ij}+{\vec v}_{0,i}+\sqrt{D_0}{\boldsymbol \xi}_i(t), 
\end{eqnarray}
where, $i,j=1, \dots N$ and $\vec{r_i}\equiv (x_i,y_i)$ defines position of the center of mass. Here, ${\vec u}_i \equiv (\partial_y, -\partial_x)\psi$ is the advection velocity and  $\vec{g} \equiv (0,-g)$ represents the drag due to particles' apparent weight~\cite{weight} (i.e., weight minus buoyant force). Further, $\vec{F}_{ij}$ represents short-ranged repulsive force derived from the LJ pair potential. 
The orientation of the self-propulsion velocity, ${\vec v}_{0,i} = v_0(\cos\theta_i,  \sin\theta_i)$ evolves as, $\dot{\theta}_i =\Omega_{0,i} + (\alpha/2)\nabla \times {\vec u}_{i}+\sqrt{D_\theta}~\xi^\theta_i(t)$. Here, $\Omega_{0,i}$ represents intrinsic torque. The  shear  torque on the AJPs is proportional to the local fluid vorticity, $\nabla \times {\vec u}_{i}$ \cite{RR2,Neufeld}. We adopt Fax\'en's second law, which, for an ideal no-stick spherical particle, yields $\alpha = 1$. 
The thermal translational noises, ${\boldsymbol
\xi}_i(t)=(\xi_{x,i}(t), \xi_{y,i}(t))$ and the rotational noise $\xi_{\theta,i} (t)$ are assumed to be
stationary, independent, delta-correlated Gaussian noises, $\langle
\xi_{\mu,i}(t)\xi_{\nu,j}(0)\rangle = 2 \delta_{ij}\delta_{\mu,\nu}\delta
(t)$, with $\mu,\nu=x,y,\theta$. The persistence time and length of self-propulsion velocity are given by, $\tau_\theta = 1/D_\theta$ and $l_\theta = \tau_\theta v_0$, respectively.
  For an AJP the rotational diffusion largely depends on the self-propulsion mechanisms. Thus, $D_0$, $v_0$, and $D_\theta$ can safely be treated as independent parameters~\cite{cataly2,Golestanian-1,Teeffelen}.

By conveniently rescaling position  $(x,y) \to (\tilde x, \tilde y)=(2\pi/L)(x,y)$ and time $t \to \tilde
t= \Omega_L t$, the tunable model parameters read
$\sigma \to (2\pi/L)\sigma$, $v_e \to v_e/U_0$, $v_0 \to
v_0/U_0$, $D_0 \to D_0/D_L$ and $D_\theta \to D_\theta/\Omega_L$.
Upon setting $L=2\pi$ and $U_0=1$, our simulation results can
be assumed in dimensionless units. However, one can easily scale back to
desired dimensional units.

The Langevin Eq.~(\ref{LE}) has been numerically integrated by means of a standard
Milstein scheme~\cite{Kloeden}. A very  short integration time step, $10^{-4} - 10^{-5}$, has been used  to ensure numerical stability.  We numerically calculate average drift velocity, fluxes at the left absorbing ($L_a$) and right  absorbing ($R_a$) ends [see Fig.1(d)] and spatial distribution of the particles. Simulation details are given in the SI (item A2).    Results reported in Fig.~2-4 are obtained by averaging over $10^3-10^4$ trajectories. We define the average drift velocity as, $ \overline{v}=\lim_{t \rightarrow \infty} (1/N) \langle  \sum_{i}[x_i(t) - x_i(0)]/t \rangle$. We express  $ \overline{v}$ in the unit of self-propulsion velocity, $\kappa = \overline{v}/v_0$. Our study focuses on the high P\'eclet  number regime where advection wins over diffusion, $D_L/D_0 \gg 1$. Further, we vary the parameters:  $g/U_0$, $ v_0/U_0 $, $\Omega_0/\Omega_L$ , and $D_\theta/\Omega_L$ over the range $0$ to $\gg1$.

{\it Autonomous AJPs ratchet ---} We first consider the limit of vanishingly small packing fraction, $\phi = 2\pi r_0^2 N/L^2 \rightarrow 0$, when AJPs can be assumed as non-interacting disks. Simulation results presented in the Fig.~2 reveal that non-interacting chiral AJPs ($|\Omega_0| > 0$)  exhibit directed autonomous motion in the convection roll arrays. The rectification power, defined as, $\eta = |\kappa| $,  can easily be  raised to {\it more than 60$\%$  by suitably adjusting advection. Further, for vanishingly small rotational diffusion {\rm [} { $ D_\theta = 0$ and  $10^{-4}$ in Fig.~2(f)} {\rm ]} the rectification power approaches to 90\%}.  The average velocity, $\overline{v}$ , is a non-monotonic function of self-propulsion parameters, $\Omega_0$ and $g$. 

\begin{figure}[tp]
\includegraphics[height=0.1\textwidth,width=0.47\textwidth]{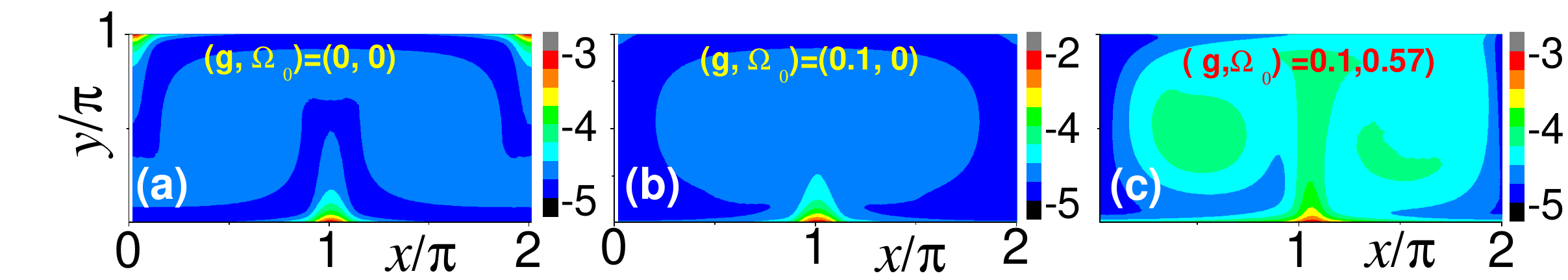}
\includegraphics[height=0.2\textwidth,width=0.45\textwidth]{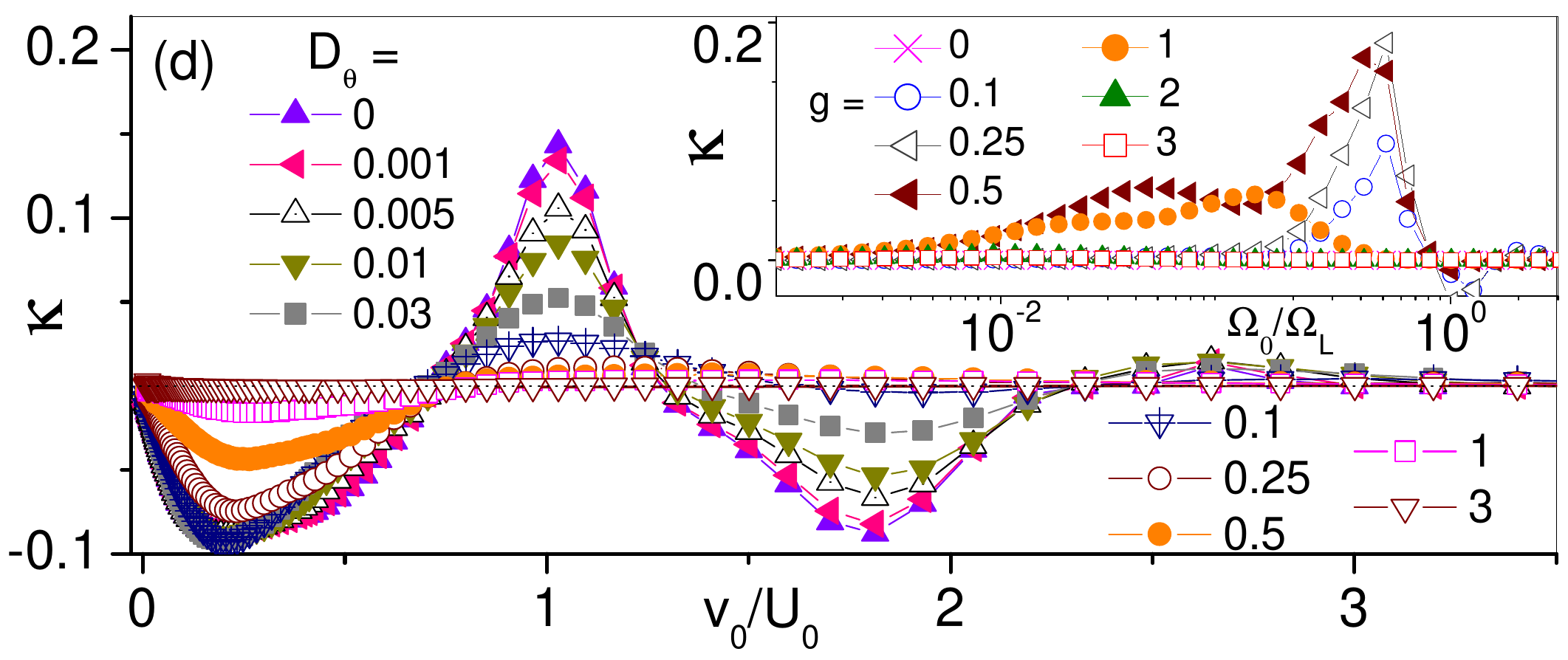}
\includegraphics[height=0.15\textwidth,width=0.45\textwidth]{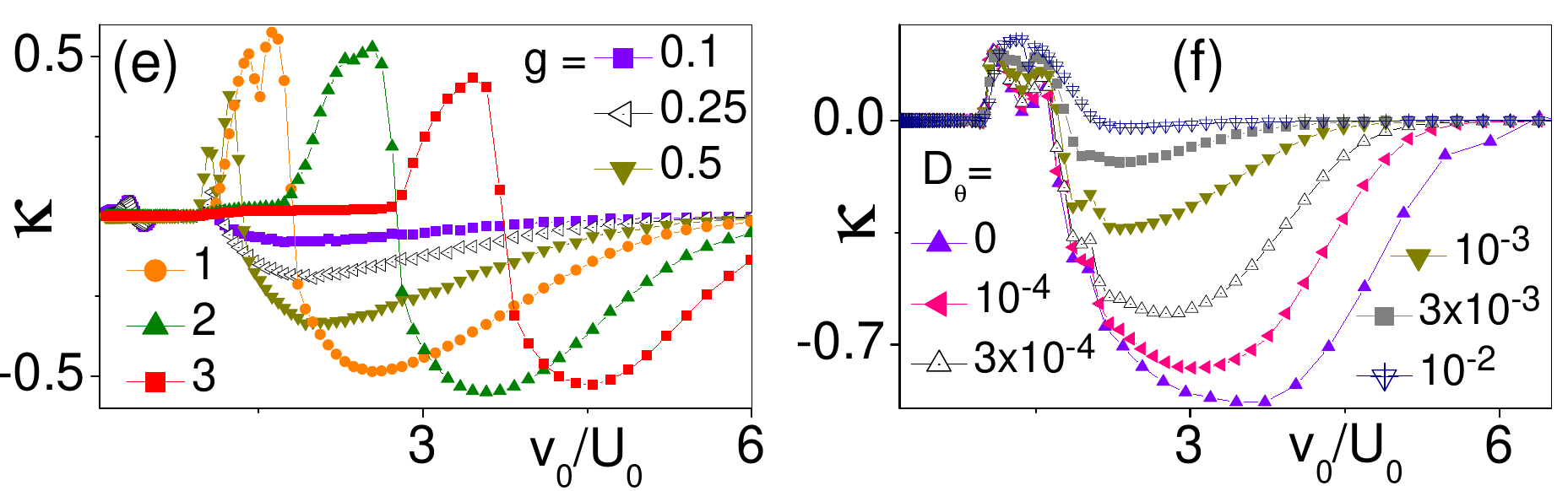}
\caption{ (Color online) (a-c) The spatial distributions, $p(x,y)$ of a non-interacting AJP in the laminar flow of Eq.~(1) for   different $\Omega_0$ and $g$ (see legends) and $ v_0= 0.2, \; D_0 = 0.01, \; D_\theta = 0.005$. The chart levels are color coded on natural logarithmic scales as indicated. (d) $\kappa \; vs. \; v_0/U_0$  for different $D_\theta$ (see legends) and $D_0 = 0.01, \; g = 0.1, \; \Omega_0 = 0.57 $. Inset: $\kappa \; vs. \; \Omega_0/\Omega_L $ with varying $g$ (see legends) and  $v_0 = 1,\; D_0 = D_\theta = 0.01$. (e) $\kappa \; vs. \; v_0/U_0$ for different $g$ (see legends) and $D_0 = 0.01, \; \Omega_0 = 0.03, \; D_\theta = 0.001$. (f) Similar plots as (d), but $D_0=0.001,\; g=1, \Omega_0 = 0.09$.}\label{F2}
\end{figure}

The symmetry breaking mechanism leading to directed autonomous motion can be understood by examining spatial distribution, $p(x,y)$ of AJPs in the convection roll arrays.  Numerical results presented in Fig.~2(a-c) [also Figure S3 of supporting information (SI)] show that when the self-propulsion length is larger than the array width, $l_\theta >L/2$,  particles tend to slide along array edges until they accumulate in the stable stagnation regions, namely at the center of the ascending (against the upper wall) and descending flows (against the lower wall). As long as,  $g\ll \{U_0 , v_0\}$, $p(x,y)$ of an achiral particle  develops a  periodic structure with apparent inversion symmetry [see Fig.~2(a)]. With growing $g$, the effective spatial structure or potential, a heavy achiral AJP experiences, loses its upside-down symmetry, however, the left-right symmetry retains there [see Fig.~2(b)].  On the other hand, a chiral AJP exhibiting orbiting motion (with radius of curvature, $R_\Omega = v_0/\Omega_0$) gets dissimilar forces in upper and lower halves of the rolls. Thus, they encounter an effective periodic structure which lacks both the left-right, as well as, the upside-down symmetry. 

Due to left-right symmetry in $p(x,y)$ of  achiral APs, the direction of rectification gets reverted on reverting intrinsic torque,  $\overline{v}(-\Omega_0) = -\overline{v}(\Omega_0)$.
However, the interplay between the advection, self-propulsion and apparent weight governs dynamical details of the rectification mechanism; hence the amplitude and direction $\overline{v}$.

In Fig.~2(d), we report $\overline{v}$ versus $v_0$ when particle's  weight is one-tenth of the maximum advection velocity  and $\Omega_0$ is comparable to the average advection torque $2\Omega_L/\pi$. For $v_0 \ll g$, AJP motion does not get rectified as self-propulsion fails to depin  particles from the stagnation areas. Again, in the opposite limit, $v_0 \gg \{U_0,\; g\}$, the particles feel  very weak {\it ratchet potential} where they exhibit free diffusion~\cite{pkg1,pkg2}, $D_{eff}=v_0^2/2D_\theta$ with   $\overline{v}_0=0$. Multiple current reversals are witnessed with the variation of $v_0$ over the range: $v_0 \ll \{g, U_0\}$ to $v_0 \gg \{g, U_0\}$.

A heavy active particle preferably falling down around descending flow center, gets advected along bottom array edges until self-propulsion is opposed by advection or the vertical component of self-propulsion surpasses the apparent weight, $v_0 \sin{\theta} > g$. Simulation results (see Fig.~2(a-c), also Figure S3-S5 in SI) well corroborate this assertion.  In the complete noiseless situation, particles are pinned at a distance, $\delta x \sim \sin^{-1}(U_0/v_0)$, from centres of the ascending flow. Here, it is assumed that no appreciable $\vec{v}_0$ direction change occurs before reaching the stagnation point. When, $v_0 \ll U_0 $,  chiral APs are pinned close to array separatrices, thus, self-propulsion helps lifting the particle along the ascending flow. 

\begin{figure}[tp]
\includegraphics[height=0.135\textwidth,width=0.45\textwidth]{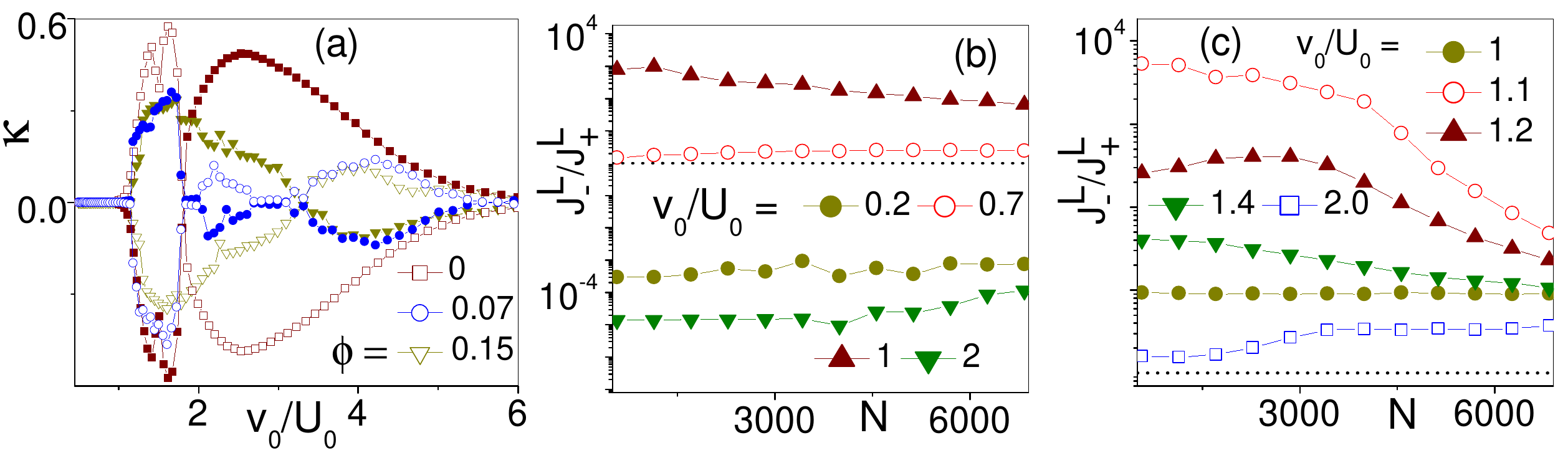}
\includegraphics[height=0.135\textwidth,width=0.45\textwidth]{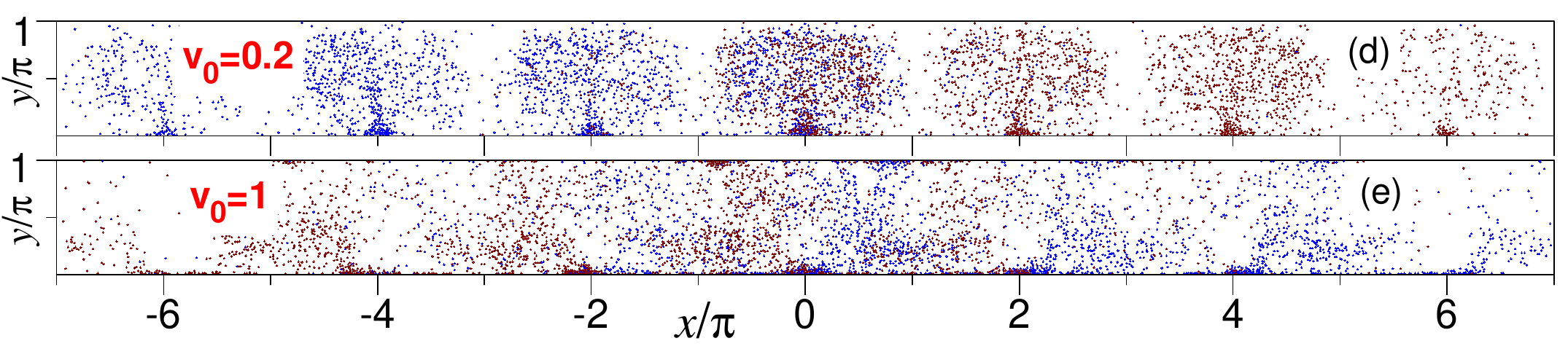}
\includegraphics[height=0.135\textwidth,width=0.45\textwidth]{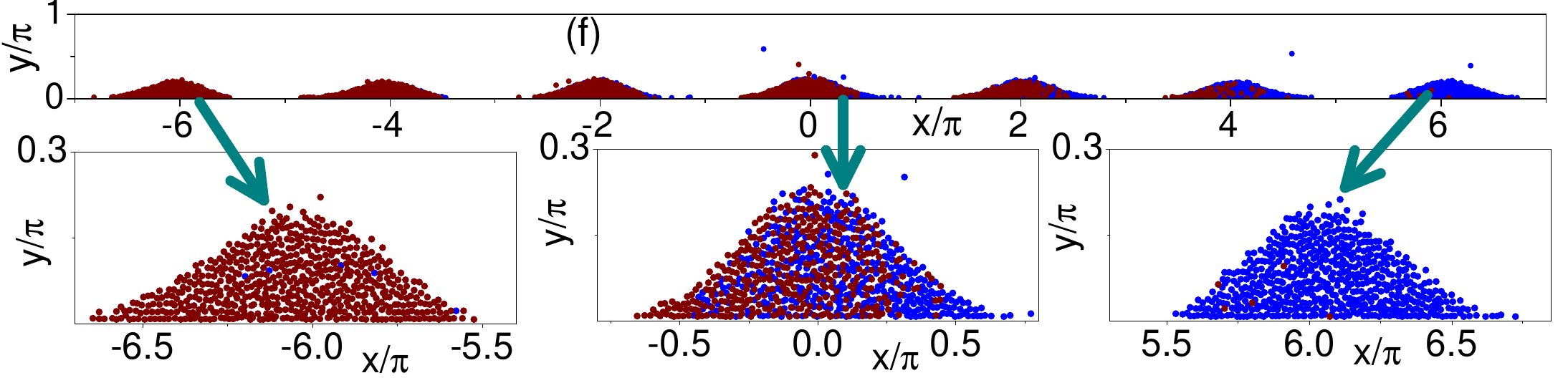}
\caption{ (Color online) (a) $\kappa \; vs. \; v_0/U_0$  for $l$ (empty symbols) and $d$ (solid symbols) AJPs in  equimolar binary mixtures  with varying $\phi$ (see legends).  Other parameters: $g =1,\; D_{\theta} = 0.001,\; D_{0} = 0.01, \Omega_0 = 0.03,\; r_0 = 0.025$. (b-c) Flux ratio of $d$ and $l$ type particles at the left-absorbing point, $J_{-}^L/J_{+}^L$  {\it vs.} $N$ for different $v_0/U_0$ (see legends); Other parameters: in (b) $g=0.25,\; D_\theta = D_0=0.01,\; r_0 = 0.025,\; \Omega_0 = 0.57 $; in (c) parameter are the same as (b), however, $g=1,\;  \Omega_0 = 0.03 $.  (d-e) Spatial distribution of equimolar binary mixture of $l$ (blue) and $d$ (wine) AJPs ($N=4560$) in the laminar flow of Eq.~(1) at $t=10^4$  starting with uniform distribution in the middle cell spanning over the range: $-\pi \; {\rm  to} \; +\pi$. Parameters having the same values as (b) but $v_0$ values are in the legends. (f) Similar plots as (d-e)  but  
 $g=1.0, \; |\Omega_0| = 0.03$ and $v_0 = 0.9$. For better visibility, the piles at the terminals  and middle cells have been zoomed in.}\label{F3}
\end{figure}

 In the upper-half of the roll, the intrinsic torque pushes the $l$ ($d$) particles to the left (right) leading to negative (positive) current. The amplitude of current keeps growing with $v_0$ as it facilitates lifting particles  against their weight. A negative current peak appears around $v_0 \sim 2g$.  Around this value of $v_0$, rectification does not require movement of AJPs through the rolls interior, thus, $\kappa$ becomes insensitive to the advection torque  (see Figure S6 of SI).

On increasing $v_0$ beyond the first peak [see Fig.~2(d)], the stagnation point shifts appreciably far from the ascending flow centres. Getting displaced there vertically, particles encounter advection torque which tends to oppose counter-clockwise rotation due to the intrinsic torque leading to decrease of negative current amplitude. With growing $\delta x$, current direction is reverted  and thereafter gets maximized around the point $v_0 \sim U_0$. For $v_0 > U_0$,  peaks  and reversal points appear around $v_0 = m\pi \Omega_0/4$, with $m=4,5,6,7 ...$.

The structure of $\kappa $ versus $v_0$ remains insensitive to the variation of both the rotational and translational diffusion constants [see Fig.~2(d) and Figure S6(b) in SI]. However, rectification power considerably enhances with lowering noise level in the system.  On the other hand, as expected,  ratcheting  does not occur for the very high noise level [$D_0 \gg D_L $ and $D_{\theta} \gg \Omega_L $].

Inset of 2(d) depicts  $\kappa$  vs.  $\Omega_0$ for different apparent weight $g$. Here, we set $v_0/U_0=1$, where current is maximum [see Fig.~2(d)]. As long as $g$ is small enough in comparison to the $v_0$ and $U_0$ prominent current peaks are observed at $\Omega_0 \sim 2\Omega_L/\pi$. However, with increasing $g$ rectification gets suppressed and peaks are shifted to the lower value of $\Omega_0$. To identify parameter regimes for better rectification power, we explore  $\kappa \; vs. \; v_0$  [see Fig.~2(e,f) with varying $g$ and $D_\theta$ for $ \Omega_0 \ll \Omega_L $] and also $\kappa \; vs. \; \Omega_0$ [shown in SI: Figure S7]. Figure 2(e,f) show that:(i) rectification becomes noticeable for $v_0 \geq g$, (ii) direction of current can be reverted by varying $v_0$ for very slow rotational diffusion, (iii) the negative current peak gets broader with lowering rotational noise strength. The rectification power around the negative current peak is in excess of $60\%$  which is much larger than the previously reported directed autonomous motion of APs. Also, self-propulsion parameters chosen here in rescaled units, are consistent with the corresponding values reported in ref.~\cite{Volpe}.  To this end, we conclude that to rectify motion of AJPs with low torque, $\Omega_0 \ll \Omega_L$, advection should be adjusted so that, $\{g/U_0,\; v_0/U_0\} \gtrsim 1$. On the other hand, for $\Omega_0 \sim \Omega_L$, ratcheting occurs  even for $g \sim U_0/10$, over a long range of $v_0$.

To check ratcheting of interacting AJPs, in Fig.~3(a) we present $\kappa$ versus $v_0/U_0$ for a binary {\it racemic} mixture~\cite{torq}. With increasing the packing fraction, the amplitude of current as well as the shape of $k $ versus $  v_0$ change noticeably [detail analysis have been provided in the SI (item A3)]. However, rectification power is high enough for experimental demonstration of this striking effect in interacting AJPs.  In the mixture, $l$ particles drift opposite to the $d$ ones. It is apparent from Fig.~3(a) that within the limit of statistical error, the relation  $\overline{v}(\Omega_0) = - \overline{v}(-\Omega_0)$ holds for the entire range of self-propulsion. Thus, $l$ and $d$ particles can easily be separated, by injecting their  mixture in a convection roll array.

\begin{figure}[tp]
\includegraphics[height=0.135\textwidth,width=0.45\textwidth]{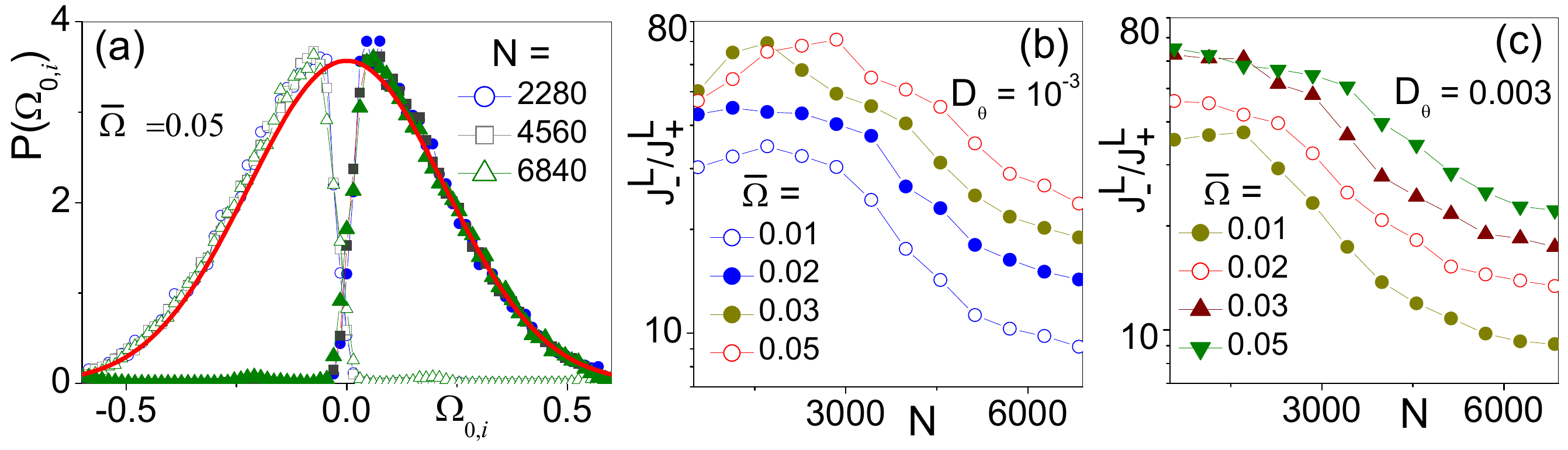}
\caption{ (Color online) (a) Torque distributions of absorbed particles in the $L_a$ (empty symbols) and $R_a$ (solid symbols) ends, when injected particle torques have Gaussian distribution (solid line) with zero mean and variance $\overline{\Omega}=0.05$. Other parameters: $D_0 = D_\theta=0.003, \; r_0 = 0.025 \; g =U_0=1, \; v_0 = 1.4$. (b,c) Ratio of $d$ and $l$ particles absorbed at the $L_a$ end for for the same parameters as (a), unless mentioned in the legends.}\label{F4} 
\end{figure}

{\it Chiral separation} --- To better illustrate the chiral separation taking advantage of ratcheting in convection roll arrays,  we consider a set up  [as shown in the Fig.~1(d)], where, particles are injected in the middle cell (spanning over $-L/2 $ to $ +L/2$) and they are absorbed/collected at $n$ cells apart $[x=\pm (k+1/2)L]$ from the injecting cell. We choose $k=3$ for all the simulation results reported in this paper. However, the conclusions of our study are insensitive to this choice. To keep particle number $N$ fixed between two absorbing points ($L_a \; {\rm and } \; R_a$),  absorbed particles are re-injected in the middle cell. We examine the distribution of particles  and flux at the absorbing points $L_a \; {\rm and } \; R_a$. First, we consider chiral separation in a {\it racemic} binary mixture~\cite{torq} where $l$ and $d$ AJPs have sharply defined torques, $+\Omega_0$ and $-\Omega_0$, respectively. Then, a more realistic situation has been considered where particle torques have Gaussian distribution with zero mean and variance $\overline{\Omega}$, $P(\Omega_{0,i})=1/\sqrt{2\pi \overline{\Omega}}\exp[-\Omega_{0,i}^2/2\overline{\Omega}]$.     

{\it Racemic binary mixture} --- Figure~3(d-f) [also, in SI: Figure S8-S10] depict distribution of AJPs [1:1 {\it dl} mixture] at $t=10^4$, starting with an uniform distribution in the middle cell. It is apparent that $l$ and $d$  particles preferably diffuse in the opposite direction to each other leading to separation of two sorts of species. Our simulation results show that the segregation mechanism for the particles with large torque ($\Omega_0 \sim \Omega_L$) works better for low g ($\sim U_0/10$)  [see Fig.~3(b, d-e)]. On the other hand, better separation of chiral particles in low torque regimes requires  $g\geq  U_0$ [Fig.~3(c,f)]. For the both limits of $\Omega_0$, the {\it separating device} operates over a long range of self-propulsion.  For $v_0 \ll \{U_0, \; g\}$, similar as noted for $\phi \rightarrow 0 $,  pinning action suppresses autonomous motion of the APs. Thus, neither separation nor flux becomes noticeable. As soon as $v_0$ approaches to $g$, separation of two sorts of species into distinct phases gets appreciable. Further, for $U_0 \sim g$ and $v_0 \geqslant U_0$  phase separation occurs with the
formation of preferably one-species clusters of different size. Cluster formation is initiated by the confining action exerted by the active particles and largely depends on the self-propulsion properties, intrinsic torques and other aspects of the microscopic mixture dynamics~\cite{MP1,MP2,MP3,MP4,MP55}  

We numerically estimate particle fluxes at the both absorbing ends, $L_a$ and $R_a$ [see Fig.1(d)].  Fluxes of $l$ and $d$  AJPs at $L_a$ end are denoted as, $J_{+}^L$  and  $J_{-}^L$, respectively. They are defined as, $J_{\pm}^{L} =  \langle n_{\pm}^{L} \rangle /t$. Where, $n_{+}^{L}$ is the number of $l$ JP absorbed at the $L_a$ end over the time $t$.  Similar notations, $J_{\pm}^{R}$, are used for fluxes at $R_a$ end. 

As current gets reverted on reverting direction of the intrinsic torques, $J_{+}^L =J_{-}^R $  and  $J_{-}^L =J_{+}^R $. The separation efficiency \cite{efficiency} can best be quantified by analysing the ratio of $l$ and $d$  particles absorbed at the $L_a$ or $R_a$ ends. Figure 3(b) and 3(c) depict variation of flux ratio, $J_{-}^L/J_{+}^L$ as a function of particle number $N$ for the high  and low torque regimes, respectively.  
 These results prove that for the both torque regimes and over a long range of  $v_0/U_0$,  almost 98\% - 99.99\% of either $l$ or $d$  AJPs are selectively absorbed at the one end. As anticipated, the segregation efficiency is much better for low density. { Simulation results show that impacts of inter-particle interactions are noticeable even for low packing fractions (for details, see item A3  of SI)}.  Further, this chiral separation {\it device} can be operated over a large parameter regimes (see SI, item A5).      
 
{\it Mixture with distributed particle torque --- } Finally, our analysis focuses on separation of $d$ and  $l$ AJPs from a mixture where particle torques are normally distributed with variance $\overline{\Omega}$.   Particle torques due to fabrication defects  cannot be considered sharply defined two values $\pm \Omega_0$ . Rather, they should be distributed over a range. 
  Figure 4(a) shows the torque distribution  of the particles absorbed in the $L_a$ and $R_a$ ends, when AJPs with Gaussian torque distribution are injected at the middle cell.  It is apparent that $d$ type AJPs are preferably  absorbed in the $L_a$ end and vice-versa. The flux ratio, $J_{-}^L/J_{+}^L$ (see Fig.4(b-c), also, Figure S11 in SI), further confirms that one can separate $d$ and  $l$ AJPs from a mixture with high efficiency. Almost, 90\% to 98.5\% separation efficiency can easily be achieved over a large  range of $\overline{\Omega},\; g\;$ and $v_0$. Strikingly, confining action initiates motility induced phase separation, thereby the $l$ and $d$ AJPs tend to pile up in the well-separated stagnation areas (shown in Figure S12-S13 of SI).   

In conclusions, our study offers an experimentally adoptable demonstration of ratcheting, as well as, an efficient way of sorting $d$ and $l$ AJPs from their mixture. { Our simulation results prove that the chiral separation device can be operated on any scale, however, it requires the ratios,  $v_0/U_0$ and $\Omega_0/\Omega_L$, to be within the range 0.1 to 2 and 0.001 to 1.0, respectively.} Further, these effects can be reproduced in artificial swimmers regardless of their propulsion mechanism, especially, not only for fabricated active Janus particles. We expect, our proposed rectification and chiral separation protocols carry huge application potentials in the micro-swimmer technology and biomedical sciences. 

\section*{Supporting Information}
The Supporting Information is available free of charge at
\url{https://pubs.acs.org/doi/10.1021/acs.jpclett.2c03193}.\\
 
 A1: Meaning of some technical terms; A2: Simulation methods; A3: Impact of inter-particle interactions on the rectification and separation; A4: Simulation results supporting the rectification mechanism; A5: Additional figures for current and flux, and spatial distribution.


\section*{Acknowledgments}
P.B. thanks UGC, New Delhi, India, for the award of a Junior
Research Fellowship. P.K.G. is supported by SERB Core Research Grant No. CRG/2021/007394.


\section*{References}
\begin{enumerate} 
%
\bibitem{Schweitzer-1} Schweitzer, F. {\it Brownian Agents and Active Particles;} Springer: Berlin, 2003.

\bibitem{Schimansky-Geier-1} Romanczuk, P.; B\"{a}r, M.; Ebeling, W.; Lindner, B.; Schimansky-Geier, L. Active Brownian Particles. From Individual to Collective Stochastic Dynamics. {\it Eur. Phys. J. Special Topics} {\bf 2012}, {\it 202}, 1-162. 

\bibitem{Ramaswamy} Ramaswamy, S. The Mechanics and Statistics of Active Matter. {\it Annu. Rev. Condens. Matter Phys.} {\bf 2010}, {\it 1}, 323-345.

\bibitem{Medin} Medina-S\'{a}nchez, M.; Schwarz, L.; Meyer, A. K.; Hebenstreit, F.; Schmidt, O. G. Cellular Cargo Delivery: Toward Assisted Fertilization by Sperm-Carrying Micromotors. {\it Nano Lett.} {\bf 2016}, {\it 16}, 555-561. 

\bibitem{JPCL-1} Wittmann, M.; Ali, A.; Gemming, T.; Stavale, F.; Simmchen, J. Semiconductor-Based Microswimmers: Attention to Detail Matters.  {\it J Phys Chem Lett.} {\bf 2021}, {\it 12}, 9651-9656.

\bibitem{Schauer} Schauer, O.; Mostaghaci, B.; Colin, R.; H\"{u}rtgen, D.; Kraus, D.; Sitti, M.; Sourjik, V. Motility and Chemotaxis of Bacteria-Driven Microswimmers Fabricated using Antigen 43-Mediated Biotin Display. {\it Sci. Rep.} {\bf 2018}, {\it 8}, 9801.

\bibitem{Magdanz} Magdanz, V.; Sanchez, S.; Schmidt, O. G. Development of a Sperm-Flagella Driven Micro-Bio-Robot. {\it Adv. Mater.} {\bf 2013}, {\it 25}, 6581-6588. 


\bibitem{JPCL-2} Holterhoff, A. L.; Li, M.; Gibbs, J. G. Self-Phoretic Microswimmers Propel at Speeds Dependent upon an Adjacent Surface's Physicochemical Properties. {\it J Phys Chem Lett.}  {\bf 2018}, {\it 9}, 5023-5028.


\bibitem{Bunea} Bunea, A.-I.; Taboryski, R. Recent Advances in Microswimmers for Biomedical Applications.  {\it Micromachines} {\bf 2020}, {\it 11}, 1048. 

\bibitem{Srivastava} Srivastava, S. K.; Medina-S\'{a}nchez, M.; Koch, B.; Schmidt, O. G. Medibots: Dual-Action Biogenic Microdaggers for Single-Cell Surgery and Drug Release. {\it Adv. Mater.} {\bf 2016}, {\it 28}, 832-837.

\bibitem{Granick} Jiang, S.; Granick, S. (Eds.), {\it Janus Particle Synthesis, Self-Assembly and Applications}; RSC Publishing: Cambridge, 2012.
%

\bibitem{JPCL-3} Tsyrenova, A.; Farooq, M. Q.; Anthony, S. M.; Mollaeian, K.; Li, Y.; Liu, F.; Miller, K.; Ren, J.; Anderson, J. L.; Jiang, S. Unique Orientation of the Solid-Solid Interface at the Janus Particle Boundary Induced by Ionic Liquids. {\it J Phys Chem Lett.}  {\bf 2020}, {\it 11}, 9834-9841.
%
\bibitem{Muller} Walther, A.; M\"{u}ller, A. H. E. Janus Particles: Synthesis, Self-Assembly, Physical Properties, and Applications. {\it Chem. Rev.} {\bf 2013}, {\it 113}, 5194-5261. 

\bibitem{Jiang} Jiang, H.-R.; Yoshinaga, N.; Sano, M. Active Motion of a Janus Particle by Self-Thermophoresis in a Defocused Laser Beam. {\it Phys. Rev. Lett.} {\bf 2010}, {\it 105}, 268302.

\bibitem{Golestanian-1} Howse, J. R.; Jones, R. A. L.; Ryan, A. J.; Gough, T.; Vafabakhsh, R.; Golestanian, R. Self-Motile Colloidal Particles: From Directed Propulsion to Random Walk. {\it Phys. Rev. Lett.} {\bf 2007},  {\it 99}, 048102.

\bibitem{Golestanian-2} Golestanian, R. Anomalous Diffusion of Symmetric and Asymmetric Active Colloids. {\it Phys. Rev. Lett.} {\bf 2009},  {\it 102}, 188305.

\bibitem{Teeffelen} Teeffelen, S. van;  L\"{o}wen, H. Dynamics of a Brownian Circle Swimmer. {\it Phys. Rev.  E} {\bf 2008}, {\it 78}, 020101(R). 

\bibitem{Stark} Z\"ottl, A.; Stark, H. Emergent Behavior in Active Colloids. {\it J. Phys.: Condens. Matter} {\bf 2016}, {\it 28}, 253001.

\bibitem{Volpe} Volpe, G.; Buttinoni, I.; Vogt, D.; K\"{u}mmerer, H.-J.; Bechinger, C. Microswimmers in Patterned Environments. {\it Soft Matter} {\bf 2011}, {\it 7}, 8810-8815.


\bibitem{Marchetti1} Fily, Y.; Marchetti, M. C. Athermal Phase Separation of Self-Propelled Particles with No Alignment.
{\it Phys. Rev. Lett.} {\bf 2012}, {\it 108}, 235702.

\bibitem{Marchetti2} Yang, X.; Manning, M. N.; Marchetti, M. C. Aggregation and Segregation of Confined Active Particles.
{\it Soft Matter} {\bf 2014}, {\it 10}, 6477-6484.



\bibitem{Redner} Redner, G. S.;  Hagan, M. F.; Baskaran, A. Structure and Dynamics of a Phase-Separating Active Colloidal Fluid. {\it Phys. Rev. Lett.} {\bf 2013}, {\it 110}, 055701.



\bibitem{Buttinoni} Buttinoni, I.; Bialk\'{e}, J.; K\"{u}mmel, F.; L\"{o}wen, H.; Bechinger, C.; Speck, T. Dynamical Clustering and Phase Separation in Suspensions of Self-Propelled Colloidal Particles.  {\it Phys. Rev. Lett.} {\bf 2013}, {\it 110}, 238301.

\bibitem{ourPRL} Ghosh, P. K.; Misko, V. R.; Marchesoni, F.; Nori, F. Self-Propelled Janus Particles in a Ratchet: Numerical Simulations. {\it Phys. Rev. Lett.} {\bf 2013}, {\it 110}, 268301. 



\bibitem{Reichhardt} Olson Reichhardt, C. J.; Reichhardt, C. Ratchet Effects in Active Matter Systems. {\it Annu. Rev. Condens. Matter Phys.} {\bf 2017}, {\it 8}, 51-75.


\bibitem{Bao} Ai, B.; Chen, Q.; He, Y.; Li, F.; Zhong, W. Rectification and Diffusion of Self-Propelled Particles in a Two-Dimensional Corrugated Channel. {\it Phys. Rev. E} {\bf 2013}, {\it 88}, 062129. 
%
\bibitem{Misko-1} Wang, X.; Baraban, L.; Nguyen, A.; Ge, J.; Misko, V. R.; Tempere, J.; Nori, F.; Formanek, P.; Huang, T.; Cuniberti, G.; Fassbender, J.; Makarov, D. High-Motility Visible Light-Driven Ag/AgCl Janus Micromotors. {\it Small} {\bf 2018}, {\it 14}, 1803613.
%
\bibitem{Jaideep} Katuri, J.; Caballero, D.; Voituriez, R.; Samitier, J.; Sanchez, S. Directed Flow of Micromotors through Alignment Interactions with Micropatterned Ratchets. {\it ACS Nano} {\bf 2018}, {\it 12}, 7282-7291.


\bibitem{Pietzonka} Pietzonka, P.; Fodor, \'{E.}; Lohrmann, C.; Cates, M. E.; Seifert, U. Autonomous Engines Driven by Active Matter: Energetics and Design Principles. {\it Phys. Rev. X} {\bf 2019}, {\it 9}, 041032.



\bibitem{soft-sepa} Mijalkov, M.; Volpe, G. Sorting of Chiral Microswimmers. {\it Soft matter} {\bf 2013}, {\it 9}, 6376.


\bibitem{our-chiral} Li, Y.; Ghosh, P. K.; Marchesoni, F.; Li, B. Manipulating Chiral Microswimmers in a Channel. {\it Phys. Rev. E} {\bf 2014}, {\it 90}, 062301.

\bibitem{cos} Costanzo, A.; Elgeti, J.; Auth, T.; Gompper, G.; Ripoll, M. Motility-Sorting of Self-Propelled Particles in Microchannels. {\it EPL} {\bf 2014}, {\it 107}, 36003.


\bibitem{der} Derivaux, J.-F.; Jack, R. L.; Cates, M. E.  Rectification in a Mixture of Active and Passive Particles subject to a Ratchet Potential. {\it J. Stat. Mech.} {\bf 2022}, {\it 2022}, 043203.

\bibitem{technical} Detailed meaning of ratchet transport and chirality have been explained in the item A1 of supporting information. 






\bibitem{torq} Particles having intrinsic torque  are referred as chiral ones. Here, AJPs with +ve and -ve torques are known as levogyre ($l$) and dextrogyre ($d$), respectively. An equal (1:1) mixture of  $l$ and  $d$  particles with intrinsic torque, $+\Omega_0$ and $-\Omega_0$, respectively, is referred here as racemic mixture, or racemate.


\bibitem{soto} Soto, F;  Wang, J.; Ahmed, R.; Demirci, U.
Medical Micro/Nanorobots in Precision Medicine, {\it Adv. Sci.} {\bf 2020}, {\it 7}, 2002203.

\bibitem{Minfeng} Zhou, M.; Ting, H.; Li, J.; Yu, S.;  Xu, Z.;  Yin, M.;  Wang, J.;  Wang, X. Self-Propelled and Targeted Drug Delivery of Poly(aspartic acid)/Iron-Zinc Microrocket in the Stomach, {\it ACS Nano } {\bf 2019}, {\it 13}, 1324 - 1332. 


\bibitem{Zhiyong} Sun, Z.;  Popp, P. F.;  Loderer, C.;  Revilla-Guarinos, A.
Genetically Engineered Bacterial Biohybrid Microswimmers for Sensing Applications, 
{\it Sensors} {\bf 2020}, {\it 20}, 180; 


\bibitem{Zoettl} Z\"{o}ttl, A.; Stark, H. Nonlinear Dynamics of a Microswimmer in Poiseuille Flow. {\it Phys. Rev. Lett.} {\bf 2012}, {\it 108}, 218104.

\bibitem{Rusconi}  Rusconi, R.; Guasto, J. S.; Stocker, R. Bacterial Transport Suppressed by Fluid Shear. {\it Nat. Phys.} {\bf 2014}, {\it 10}, 212-217.

\bibitem{Qi} Qi, K.; Annepu, H.; Gompper, G.; Winkler, R. G. Rheotaxis of Spheroidal Squirmers in Microchannel Flow: Interplay of Shape, Hydrodynamics, Active Stress, and Thermal Fluctuations. {\it Phys. Rev. Res.} {\bf 2020}, {\it 2}, 033275.


\bibitem{Kirby} Kirby, B. J. {\it Micro- and Nanoscale Fluid Mechanics: Transport in Microfluidic Devices}; Cambridge University Press, 2010.

\bibitem{Bodenschatz} Bodenschatz, E.; Pesch, W.; Ahlers, G. Recent Developments in Rayleigh-B\'{e}nard Convection. {\it Annu. Rev. Fluid Mech.} {\bf 2000}, {\it  32}, 709-778.

\bibitem{Getling} Getling, A. V. {\it Rayleigh-B\'{e}nard Convection: Structures and Dynamics}; World Scientific: Singapore, 1998.

\bibitem{wan1} Wang, X.; Liu, M.; Jing, D.; Mohamad, A.; Prezhdo. O.  Net Unidirectional Fluid Transport in Locally Heated Nanochannel by Thermo-osmosis. {\it Nano Lett.} {\bf 2020}, {\it 20}, 8965-8971.

\bibitem{wan2} Wang, X.; Liu, M. Jing, D.; Prezhdo, O.  Generating Shear Flows without Moving Parts by Thermo-osmosis in Heterogeneous Nanochannels. {\it J. Phys. Chem. Lett.} {\bf 2021}, {\it 12}, 10099-10105.

\bibitem{wan3} Wang, X.; Jing, D.  Directional Manipulation of Diffusio-Osmosis Flow by Design of Solute-Wall and Solvent-Wall Interactions. {\it J. Phys. D: Appl. Phys.} {\bf 2022}, {\it 55}, 145401.

\bibitem{hul}  Hulme, S. E.; Shevkoplyasa, S. S.; Whitesides, G. M. Incorporation of Prefabricated Screw, Pneumatic, and Solenoid Valves into Microfluidic Devices. {\it Lab Chip} {\bf 2009}, {\it 9},  79-86.
    
\bibitem{Ran} Lei, Q.-L.; Ciamarra, M. P.; Ni, R. Nonequilibrium Strongly Hyperuniform Fluids of Circle Active Particles with Large Local Density Fluctuations. {\it Sci. Adv.} {\bf 2019}, {\it 5}, eaau7423.

  
\bibitem{weight} Even it could be considered as a force due to some external field.

\bibitem{RR2} Li, Y.; Li, L.; Marchesoni, F.; Debnath, D.; Ghosh, P. K. Diffusion of Chiral Janus Particles in Convection Rolls. {\it Phys. Rev. Res.} {\bf 2020}, {\it 2}, 013250.
%
\bibitem{Neufeld} Torney, C.; Neufeld, Z. Transport and Aggregation of Self-Propelled Particles in Fluid Flows. {\it Phys. Rev. Lett.} {\bf 2007}, {\it 99}, 078101.
%
\bibitem{cataly2} Gibbs, J. G.; Zhao, Y.-P. Autonomously Motile Catalytic Nanomotors by Bubble Propulsion. {\it Appl. Phys. Lett.} {\bf 2009}, {\it 94}, 163104.

\bibitem{Kloeden} Kloeden, P. E.; Platen, E. {\it Numerical Solution of Stochastic Differential Equations}; Springer: Berlin, 1992.

\bibitem{pkg1} Ghosh, P. K.; Marchesoni, F.; Li. Y.; Nori, F.  Active Particle Diffusion in Convection Roll Arrays. {\it Phys. Chem. Chem. Phys.} {\bf 2021}, {\it 23}, 11944-11953.

\bibitem{pkg2} Ghosh, P. K.; Debnath, D.; Li, Y. Marchesoni, F.  Diffusion of Active Particles in Convective Flows. {\it Soft Matter} {\bf 2021}, {\it 17}, 2256-2264.
    


 

%
\bibitem{MP1} Matas-Navarro, R.; Golestanian, R.; Liverpool, T. B.; Fielding, S. M. Hydrodynamic Suppression of Phase Separation in Active Suspensions. {\it Phys. Rev. E} {\bf 2014}, {\it 90},  032304.


\bibitem{MP2} Theers, M.; Westphal, E.; Qi, K.; Winkler, R. G.; Gompper, G. Clustering of Microswimmers: Interplay of Shape and Hydrodynamics. {\it Soft Matter} {\bf 2018}, {\it 14}, 8590. 

\bibitem{MP3} St\"urmer, J.; Seyrich, M.; Stark, H. Chemotaxis in a Binary Mixture of Active and Passive Particles. {\it J. Chem. Phys.} {\bf 2019}, {\it 150}, 214901.
%
\bibitem{MP4}  Agudo-Canalejo, J.; Golestanian, R. Active Phase Separation in Mixtures of Chemically Interacting Particles. {\it Phys. Rev. Lett.} {\bf 2019}, {\it 123},  018101.
%
\bibitem{MP55} Dolai, P.; Simha, A.; Mishra, S. Phase Separation in Binary Mixtures of Active and Passive Particles. {\it Soft Matter} {\bf 2018}, {\it 14},  6137.
    
\bibitem{efficiency} Separation efficiency or selectivity is defined as, $\eta_s = max(J_{+}^L,J_{-}^L)/(J_{+}^L+J_{-}^L) = max(J_{+}^R,J_{-}^R)/(J_{+}^R+J_{-}^R)$.     


\end{enumerate}
\end{document}